\documentclass[aps,prb,twocolumn,floatfix,showpacs,citeautoscript,longbibliography,hyperlinks,superscriptaddress]{revtex4-2}

\usepackage{amsmath,amssymb,amsfonts}

\usepackage{graphicx}
\usepackage{comment}

\usepackage[colorlinks=true,citecolor=blue]{hyperref}
\apptocmd{\sloppy}{\hbadness 10000\relax}{}{}

\usepackage{xcolor}

\usepackage{CJK} 

\newcommand{\revision}[1]{{{#1}}}

\begin{document}

\begin{CJK*}{UTF8}{gbsn} 

\title{Heat, work, and fluctuations in a driven quantum resonator}
\author{Riya Baruah}
\affiliation{Department of Applied Physics, Aalto University, 00076 Aalto, Finland}
\author{Pedro Portugal}
\affiliation{Department of Applied Physics, Aalto University, 00076 Aalto, Finland}
\author{Jun-Zhe Chen (陈俊哲)}
\affiliation{Department of Applied Physics, Aalto University, 00076 Aalto, Finland}
\author{Joachim Wabnig}
\affiliation{Nokia Bell Labs, Cambridge, United Kingdom}
\author{Christian Flindt}
\affiliation{Department of Applied Physics, Aalto University, 00076 Aalto, Finland}

\begin{abstract}
A central building block of a heat engine is the working fluid, which mediates the conversion of heat into work. In nanoscale heat engines, the working fluid can be a quantum system whose behavior and dynamics are non-classical. A particularly versatile realization is a quantum resonator, which allows for precise control and coupling to thermal reservoirs, making it an ideal platform for exploring quantum thermodynamic processes. Here, we investigate the thermodynamic properties of a driven quantum resonator whose  temperature is controlled by modulating its natural frequency. We evaluate the work performed by the external drive and the resulting heat flow between the resonator and its environment, both within linear response and beyond. To further elucidate these processes, we determine the full distribution of photon exchanges between the resonator and its environment, characterized by its first few cumulants. Our results provide quantitative insights into the interplay between heat, work, and fluctuations, and may help in designing future heat engines.
\end{abstract}

\maketitle
\end{CJK*}

\section{Introduction} 

Heat engines are fundamental devices for converting thermal energy into useful work, and they have played a central role in the development of thermodynamics~\cite{Callen1985}. Typically, their operation is based on a macroscopic working fluid whose behavior is well captured by classical physics. Recent advances in nanofabrication, quantum control, and measurement techniques, however, have enabled the realization of heat engines operating at the nanoscale, where the working fluid may involve only a few degrees of freedom that must be described quantum mechanically~\cite{Vinjanampathy:2016,Strasberg:2022}. In this regime, quantum coherence, fluctuations, and measurement backaction can significantly modify the thermodynamic behavior, raising intricate questions about the nature of work, heat, and efficiency beyond classical physics~\cite{Giazotto:2006,Quan:2007,Cangemi:2024}. Understanding how these quantities behave and interplay in driven quantum systems is therefore essential both for advancing quantum thermodynamics and for developing future quantum devices for efficient energy conversion.

Examples of working fluids in quantum heat engines include qubits~\cite{Ono2020, Kamimura2022}, quantum harmonic oscillators~\cite{Leitch2022, Insinga2016}, and other multilevel systems~\cite{Scovil:1959,Quan2005, Anka2021}. Experimentally, quantum thermodynamics has been explored in a variety of platforms, such as electronic circuits~\cite{Pekola2015r, Hofmann:2016}, trapped ions~\cite{An:2014,Rossnagel:2016}, and superconducting devices~\cite{Masuyama2018,Pekola:2013,Ronzani:2018,Aamir2025}. 
Unlike in macroscopic systems, where fluctuations vanish due to the sheer system size and the large number of particles involved, quantum mechanical working fluids may exhibit quantum and thermal fluctuations that are not small compared to average values. The discrete transitions in a quantum heat engine may soon be experimentally accessible thanks to developments in quantum calorimetry and the non-invasive detection of photon exchanges~\cite{Anders2013,Brange:2018, Karimi2020, KarimiPekola2020}. The stochastic nature of these transitions requires a statistical analysis, for instance, to understand fluctuations of heat and work \cite{Silaev:2014,Gasparinetti:2014},  photon-number statistics \cite{Clerk:2007,Clerk:2011,Hofer:2016}, and the distribution of photon waiting times~\cite{Brange2019,Menczel:2021, Portugal:2023,Kansanen:2025}.  These experimental and theoretical advances motivate investigations of quantum systems as the building blocks of heat engines and refrigerators. 

\begin{figure}[!b]
    \centering
\includegraphics[width=\columnwidth]{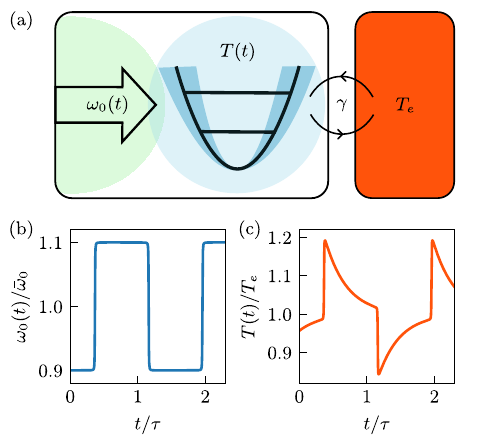}
    \caption{Driven quantum resonator. (a) An external drive performs work on the resonator by changing its natural frequency, $\omega_0(t)$,  which leads to a changing temperature, $T(t)$. The temperature of the resonator also changes because heat is exchanged with a thermal reservoir at the temperature $T_e$. The coupling to the reservoir is denoted by~$\gamma$. (b) Square-wave drive of the resonator frequency. (c) Corresponding temperature of the resonator. Parameters are $k_B T_e = 1.5 \hbar \bar \omega_0$ in terms of the resonator frequency without the drive. The period and amplitude of the drive are $\tau = 2 \pi /(0.1 \bar \omega_0)$ and $\Delta\omega_0=0.1\bar \omega_0$.}
    \label{Fig:1}
\end{figure}

\revision{While driven resonators and their thermodynamic properties have been investigated extensively, most works have focused either on average thermodynamic quantities or on stationary fluctuation properties~\cite{Leitch2022,Deffner:2008,Deffner:2010,Denzler:2018,Vasquez2018}. In contrast, our work provides a unified, time-resolved description of both average thermodynamic observables and the full counting statistics of photon transfers under arbitrary temporal frequency modulation. Specifically, our approach combines explicit time-dependent control of the resonator temperature via frequency modulation, a description of the photon counting statistics in a nonstationary setting using simple scalar equations, and a systematic comparison between nonlinear dynamics and linear-response predictions, both for average thermodynamic quantities as well as higher-order cumulants. Our framework allows us to investigate how fluctuations evolve dynamically during thermodynamic processes, going beyond stationary or cycle-averaged descriptions.}

The setup that we consider is illustrated in Fig.~\ref{Fig:1}(a): A quantum resonator with a controllable frequency, which we analyze as the potential working fluid of a quantum heat engine. The external drive of the resonator frequency performs work on the resonator, which consequently changes its temperature. At the same time, heat flows between the resonator and its thermal environment because of the temperature difference between them~\cite{portugal2022, Baruah:2025}. In Fig.~\ref{Fig:1}(b), we show an example of a square-wave drive, which causes instantaneous changes of the resonator temperature as seen in Fig.~\ref{Fig:1}(c). In between these sudden temperature changes, the resonator relaxes towards equilibrium with the reservoir. We analyze the temperature of the resonator for different driving schemes together with the injected power and the heat flow. We also investigate the statistics of photon exchanges between the resonator and the thermal environment. \revision{The main contributions of our work are: (1) A time-resolved thermodynamic description of a driven quantum resonator beyond steady-state or cycle-averaged analyses. (2) A derivation of dynamical equations for the photon counting cumulants under time-dependent driving. (3) Identification of qualitatively distinct temporal behavior of higher-order cumulants, revealing non-Gaussian fluctuation dynamics. (4) A linear-response theory for the thermodynamic observables and the high-order cumulants, showing how they encode independent dynamical information beyond mean values.} These findings may find use in the design and optimization of nanoscale heat engines based on quantum resonators as the working fluid. 

The rest of our article is organized as follows. In Sec.~\ref{sec:driven_res}, we introduce our model of a driven quantum resonator coupled to a heat reservoir. In Sec.~\ref{sec:temp}, we demonstrate how the temperature of the resonator can be controlled by modulating its natural frequency in time. In Sec.~\ref{sec:PandJ}, we show how the temperature modulations can be understood as an interplay between the power that is supplied by the external drive and the heat that is exchanged with the reservoir. In Sec.~\ref{sec:LR1}, we analyze the setup in linear response, where the driving amplitude is small, and the thermodynamic observables can be found analytically. In Sec.~\ref{sec:fluc}, we go beyond investigations of mean values by considering the distribution of photon exchanges between the resonator and its environment. In Sec.~\ref{sec:cumu}, we find the first few cumulants of the distribution, which are further analyzed in Sec.~\ref{sec:LR2}, where we consider the linear-response regime. In Sec.~\ref{sec:FCS}, we then evaluate the full distribution of photon exchanges, which are characterized by the first few cumulants from the previous sections. Finally, in Sec.~\ref{sec:concl}, we present our conclusions and provide an outlook on possible extensions of our work.

\section{Driven resonator}
\label{sec:driven_res}

The Hamiltonian of the driven resonator reads
\begin{equation}
    \hat H(t) =\hbar\omega_0(t)(\hat a^\dagger \hat a + 1/2),
\label{eq:resonatorH}
\end{equation}
where the time-dependence enters through the resonator frequency, which is modulated in time. This particular form of the Hamiltonian assumes that both the kinetic and potential energy are changed in time. For example, for an $LC$-resonator, both the inductance and the capacitance would be modulated. As such, our setup differs from the common one, where only the potential energy is modulated to change the  frequency~\cite{Deffner:2008, Deffner:2010,Denzler:2018,Vasquez2018}.
\revision{Importantly, our  Hamiltonian enables a direct mapping between frequency modulation and  temperature changes of the oscillator, cf.~Eq.~(\ref{eq:Wtemp}). This property is not generic to  driven oscillators, and it provides a transparent thermodynamic interpretation of the drive as an efficient way of controlling the temperature, which we exploit in this work.}

The resonator is weakly coupled to a thermal reservoir at the temperature $T_e$, such that its density matrix evolves according to the Lindblad master equation
\begin{equation}
\label{eq:LindbladME}
    \partial_t\hat\rho = \mathcal{L}(t) \hat \rho= \frac{1}{i\hbar} [\hat H(t), \hat \rho] + \gamma \mathcal{D}\hat \rho,
\end{equation}
where $\gamma$ is the strength of the coupling to the environment, and we have introduced the  dissipator
\begin{equation}
\begin{split}    
\mathcal{D}\hat \rho = 
\big(1&+n_B(\omega_0)\big)\big(\hat a \hat \rho \hat a^\dagger - \{\hat a^\dagger \hat a, \hat \rho\}/2\big)\\
&+n_B(\omega_0)\big(\hat a^\dagger \rho \hat a - \{\hat a \hat a^\dagger, \hat \rho\}/2\big)
\end{split}
\end{equation}
and the Bose-Einstein distribution 
\begin{equation}
\label{eq:BEdist}
n_{B}(\omega_0) = \frac{1}{e^{\hbar \omega_0/k_B T_e} - 1}.
\end{equation}

In Eq.~(\ref{eq:LindbladME}), we have also defined the Lindblad operator, $\mathcal{L}(t)$, which generates the time-evolution.
To simplify the notation, we do not always indicate the explicit time-dependence of all quantities and operators; however, throughout this work, it should be kept in mind that most quantities are time-dependent because of the time-dependent resonator frequency, $\omega_0=\omega_0(t)$. 

The derivation of the master equation can be found in Ref.~\cite{portugal2022}, and it relies on the Hamiltonian of the resonator commuting with itself at different times because of its specific form in Eq.~(\ref{eq:resonatorH}). For this reason, the resulting master equation is valid on all timescales that are longer than the correlation time of the thermal reservoir, which in practice can be assumed to be very short.

\section{Time-dependent temperature}
\label{sec:temp}

Initially, at $t=t_0$, the resonator is in thermal equilibrium with the environment, and its density matrix reads 
\begin{equation}
\label{eq:effectivethermal}
    \hat\rho(t_0)=e^{-\hat H(t_0)/k_B T(t_0)}/Z,
\end{equation} 
where
\begin{equation}
Z=\mathrm{tr}\{e^{-\hat H(t_0)/k_B T(t_0)}\}
\end{equation}
is the canonical partition function, and the initial temperature, $T(t_0)=T_e$, is given by the temperature of the environment. We then start to modulate the resonator frequency and evolve the density matrix according to the Lindblad master equation. Assuming first that the modulations are much faster than the timescale set by the coupling $\gamma$, we can ignore the time-evolution due to the dissipator and focus only on the unitary dynamics. 
The density matrix then becomes
\begin{equation}
    \hat\rho(t_1)=\hat U(t_1,t_0)\hat\rho(t_0)\hat U^\dagger(t_1,t_0)=\hat\rho(t_0),
\end{equation} 
since the time-ordered exponential of the Hamiltonian
\begin{equation}\hat U(t_1,t_0)=\hat{T}\left\{e^{-i\int_{t_0}^{t_1}dt \,\hat{H}(t)/\hbar} \right\}
\end{equation}
commutes with the initial density matrix. Thus, the density matrix does not change. Still, the Hamiltonian has changed, and we can express the density matrix as
\begin{equation}
\label{eq:rhotemp}
    \hat\rho(t_1)=e^{-\hat H(t_1)/k_B T(t_1)}/Z,
\end{equation} 
where
\begin{equation}
    T(t_1)=\frac{\omega_0(t_1)}{\omega_0(t_0)}T(t_0)
\label{eq:Wtemp}
\end{equation} 
is now the temperature of the resonator. Indeed, at the time $t=t_1$, the resonator would be in equilibrium with a thermal reservoir at the temperature $T(t_1)$. Using this approach, one could for example prepare the resonator at a different temperature than the reservoir and investigate the thermalization process as done in Ref.~\cite{Denzler:2018}.

\begin{figure*}[!t]
	\centering
	\includegraphics[width = \textwidth]{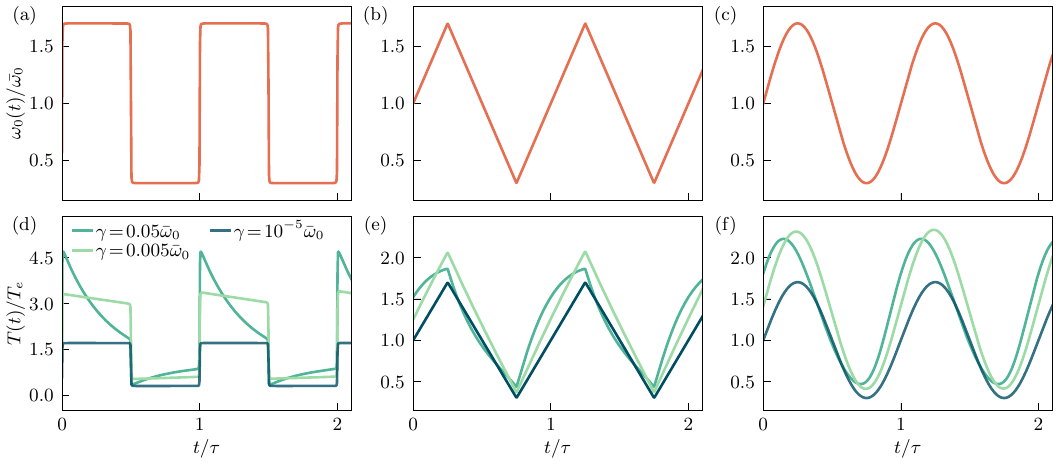}
	\caption{External drive and time-dependent temperature. (a,b,c) Square-wave, sawtooth, and harmonic drive of the resonator frequency. (d,e,f) Time-dependent temperature of the resonator for three different couplings to the environment. For the smallest coupling, the temperature follows the drive according to Eq.~(\ref{eq:Wtemp}). The reservoir temperature is $k_B T_e = 1.5 \hbar \bar \omega_0$ in terms of the resonator frequency without the drive. The period and amplitude of the drive are $\tau = 2 \pi /(0.1 \bar \omega_0)$ and $\Delta\omega_0=0.7\bar \omega_0$.} 
	\label{fig:fig2}
\end{figure*}

The temperature of the resonator changes because work is performed on it. This situation is analogous to an adiabatic compression, in which a piston reduces the volume of a gas in a cylinder and performs work against the pressure. The work increases the energy of the gas and thus its temperature. If the compression is so fast that no heat escapes, no entropy is produced and the process is reversible. By reversing the motion of the piston, the temperature can be lowered back to its original value.

We now consider the general case where heat is exchanged between the resonator and the reservoir. Importantly, if the resonator is in a thermal state, and it evolves according to the Lindblad equation, Eq.~(\ref{eq:LindbladME}), it will remain in a thermal state. However, its temperature may be different from Eq.~(\ref{eq:Wtemp}), which only holds in the absence of heat flow. Still, we can find the time-dependent temperature of the resonator. To this end, we use the population equation for the resonator
\begin{equation}
\partial_t n(\omega_0)= \gamma \big(n_B(\omega_0)-n(\omega_0)\big),
\label{eq:n(omega)}
\end{equation}
which follows from Eq.~(\ref{eq:LindbladME})~\cite{portugal2022}. Here, the Bose-Einstein distribution $n_B(\omega_0)$ is given by the temperature of the environment, while $n(\omega_0)$ is the average occupation of the resonator, expressed in terms of its temperature as
\begin{equation}
n(\omega_0) = \frac{1}{e^{\hbar \omega_0/k_B T} - 1}.
\label{eq:resn}
\end{equation}
Thus, by solving Eq.~(\ref{eq:n(omega)}) for $n(\omega_0)$, we can determine the resonator temperature from Eq.~(\ref{eq:resn}).  Moreover,
from Eq.~(\ref{eq:n(omega)}), we recover Eq.~(\ref{eq:Wtemp}) without the coupling to the environment. Indeed, for $\gamma=0$, the right-hand side vanishes, and the ratio $\omega_0(t)/T(t)$ must be constant.   

In Fig.~\ref{fig:fig2}, we show the time-dependent resonator temperature for a square-wave drive, a sawtooth drive, and a harmonic drive, with different values of the coupling to the environment. In all figures, we consider the situation, where the system has reached a periodic state, and the initial transients have decayed. For the smallest coupling, the temperature follows the drive as expected according to Eq.~(\ref{eq:Wtemp}). By contrast, as the coupling to the environment is increased, the temperature of the resonator is rather determined in an interplay between the power that is supplied by the external drive and the heat that flows between the resonator and the reservoir. For example, for the square-wave drive, the resonator begins to equilibrate with the reservoir between the fast changes of the resonator frequency. In all three cases, the temperature response to the modulated resonator frequency clearly becomes nonlinear as the coupling is increased.

\revision{In general, the behavior is governed by the competition between two timescales: the period of the drive,   $\tau$, and the relaxation time, $1/\gamma$. In the limit $\gamma\tau\ll 1$, the dynamics is effectively adiabatic in the thermodynamic sense (no heat exchange), while for $\gamma\tau> 1$, irreversible heat exchange leads to entropy production and deviations from Eq.~(\ref{eq:Wtemp}). This crossover is central to understanding the performance of driven quantum heat engines.}
	
\section{Power and heat}
\label{sec:PandJ}

To better understand the temperature modulations, we consider the average energy of the resonator measured from the ground state energy, $E_0(t)=\hbar\omega_0(t)/2$,
\begin{equation}
      U(t) = \text{tr} \big\{ \big(\hat H(t)-E_0(t)\big) \hat  \rho (t)\big \}.
\end{equation}
The change of the resonator energy then reads
\begin{equation}
     \partial_t U(t) \!=\! \text{tr} \big\{ \hat  \rho (t) \partial_t \big(\hat H(t)\!-\! E_0(t)\big)\big\}\\
     + \text{tr} \big\{\hat H(t) \partial_t \hat \rho(t)\big\},
\end{equation}
where we have used that $ \text{tr} \{\partial_t \hat \rho(t)\}=0$ because of trace conservation. We can also identify the injected power
\begin{equation}
     P(t) = \text{tr} \big\{ \hat  \rho (t) \partial_t \big(\hat H(t)-E_0(t)\big)\big\}=n(\omega_0) \partial_t(\hbar\omega_0)
\end{equation}
and the heat that flows into the resonator
\begin{equation}
    J(t) = \text{tr} \{\hat H(t) \partial_t \hat \rho(t)\}=\hbar \omega_0 \gamma\big(n_B(\omega_0)-n(\omega_0)\big).
\end{equation}

\begin{figure*}[!t]
	\centering
	\includegraphics[width = \textwidth]{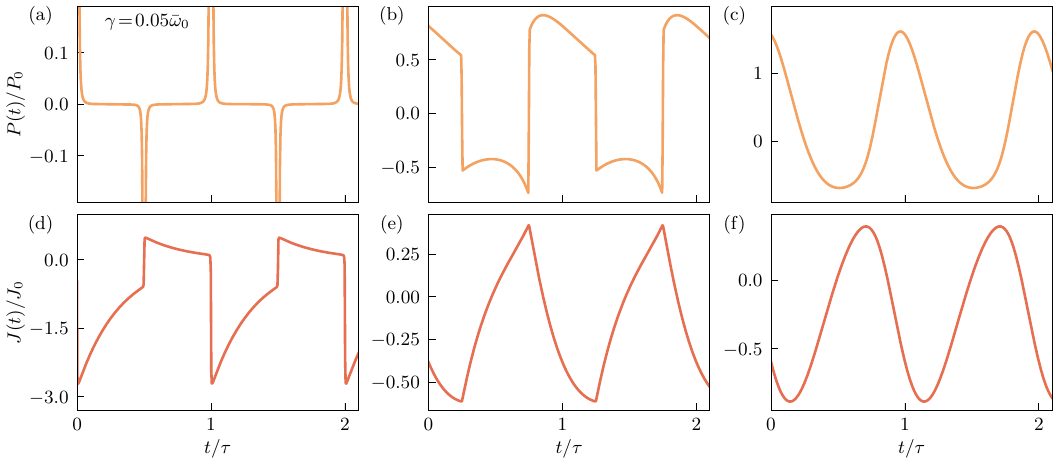}
	\caption{Power and heat. (a,b,c) Power for the threes drives in Fig.~\ref{fig:fig2}. (d,e,f) Heat currents for the three drives. The value of the reservoir coupling is $\gamma=0.05\bar\omega_0$, while the other parameters are the same as in Fig.~\ref{fig:fig2}. We have also defined $P_0=J_0=\hbar\bar\omega_0/\tau$.}
	\label{fig:fig3}
\end{figure*}

In Fig.~\ref{fig:fig3}, we show the power and the heat corresponding to the three drives in Fig.~\ref{fig:fig2}, focusing on a single value of the coupling to the environment. For the square-wave drive, the sharp pulses in the power lead to an abrupt change of the resonator temperature, which is followed by heat being transmitted into the reservoir or absorbed from it, depending on the temperature difference. A similar behavior can be observed for the two other drives. \revision{The clear separation between power pulses and subsequent heat flow highlights a temporal ordering of work and dissipation, which is a key ingredient in thermodynamic cycles. In particular, the results suggest how specific driving protocols can be engineered to maximize work extraction before dissipation sets in, which is  relevant for optimizing finite-time heat engines.}

\section{Linear response}
\label{sec:LR1}

If the amplitude of the driving is small enough, we can calculate the thermodynamic observables from above in the linear regime. To this end, we consider a small time-dependent change of the resonator frequency, 
\begin{equation}
\omega_0(t) = \bar \omega_0 + \Delta\omega_0(t),     
\label{eq:InputLR}
\end{equation}
where $\bar\omega_0$ is the frequency without the drive. The temperature, the power, and the heat then take the form
\begin{equation}
\begin{split}
T(t) &= \bar T + \Delta T(t),\\
P(t) &= \bar P + \Delta P(t),\\
J(t) &= \bar J+ \Delta J(t),     
\end{split}
\label{eq:OutputLR}
\end{equation}
where $\bar T = T_e$, $\bar P=0$, and $\bar J =0$ denote these quantities in the absence of the drive, and the response can be calculated to linear order in the perturbation, $\Delta\omega_0(t)$.

For the change of the temperature, we first expand
\begin{equation}
\begin{split}
    \Delta n_B(\omega_0)\simeq \frac{\hbar n'_B(\bar\omega_0)}{k_BT_e}\Delta\omega_0(t)
\end{split}
\end{equation}
and
\begin{equation}
    \Delta n(\omega_0)\simeq \frac{\hbar n'_B(\bar\omega_0)}{k_BT_e}(\Delta\omega_0(t)-\bar\omega_0\Delta T(t)/T_e)
\end{equation}
in the frequency and temperature variations, where 
\begin{equation}
\Delta n_B(\omega_0)=n_B(\omega_0)-n_B(\bar\omega_0)
\end{equation}
and 
\begin{equation}
\Delta n(\omega_0)=n(\omega_0)-n_B(\bar\omega_0).
\end{equation}
Inserting these expansions into Eq.~(\ref{eq:n(omega)}), we can solve for the temperature in the Fourier domain,
\begin{equation}
    \Delta T(\Omega) = \frac{i \Omega}{\gamma+i \Omega } \frac{T_e}{\bar \omega_0}  \Delta\omega_0(\Omega),
    \label{eq:TLR}
\end{equation}
where $\Omega$ is the Fourier variable (or the frequency of a harmonic drive). The power can be approximated as
\begin{equation}
        P(t) \simeq n_B(\bar \omega_0) \hbar \partial_t \Delta\omega_0(t),
\end{equation}
where we have set $n(\omega_0)=n_B(\bar\omega_0)$, so that the power is linear in the change of the resonator frequency. In the Fourier domain, we then find 
\begin{equation}
         P(\Omega) \simeq  i \Omega n_B(\bar \omega_0) \hbar \Delta \omega_0(\Omega).
         \label{eq:PLR}
 \end{equation}
Similarly, we can linearize the heat current as
\begin{equation}
    J(t) \simeq \hbar \bar\omega_0 \partial_t n(\omega_0),
\end{equation}
and in the Fourier domain, we then find
\begin{equation}
    J(\Omega) \simeq \hbar \bar\omega_0 \frac{i\gamma\Omega}{\gamma + i\Omega}n_B'(\bar\omega_0)\Delta \omega_0(\Omega),
    \label{eq:JLR}
\end{equation}
where we have first solved Eq.~(\ref{eq:n(omega)}) in the Fourier domain.

\begin{figure*}
	\centering
	\includegraphics[width = \textwidth]{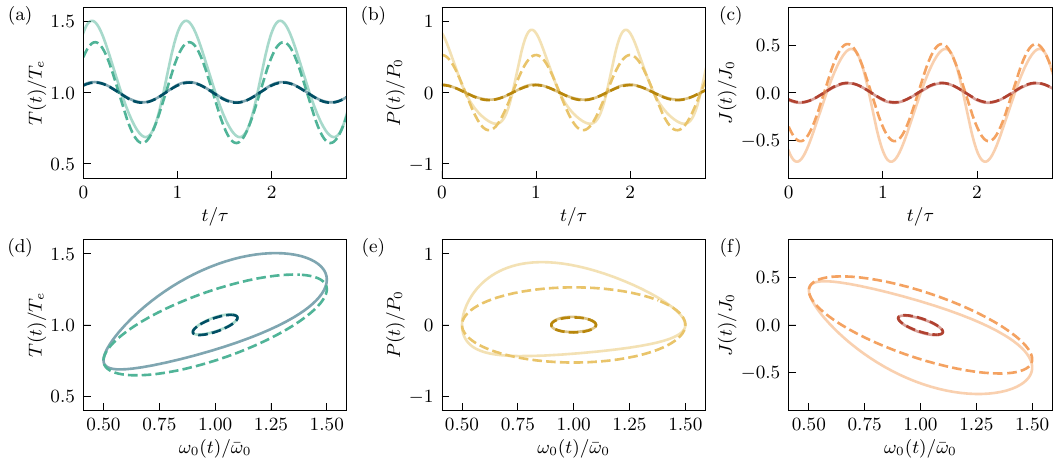}
	\caption{Linear response. (a,b,c) Temperature, power, and heat for the harmonic drive $\omega_0(t)=\bar\omega_0+\Delta\omega_0\sin(\Omega t)$ with $\Omega = 0.1\bar\omega_0$ and $\Delta \omega_0 = 0.1 \bar \omega_0, 0.5 \bar \omega_0$. The other parameters are   
		$\gamma = 0.1 \bar\omega_0$ and $k_B T_e = 1.5 \hbar \bar \omega_0$. Solid lines show numerical results, while dashed lines are the linear-response expressions in Eqs.~(\ref{eq:TLR}), (\ref{eq:PLR}), and (\ref{eq:JLR}). (d,e,f) Same quantities in a parametric plot.} 
	\label{fig:fig4}
\end{figure*}

In Fig.~\ref{fig:fig4}, we show the temperature, power, and heat for a harmonic drive with either a small or a large  amplitude. Together with the numerical results (solid lines), we show the linear-response predictions based on Eqs.~(\ref{eq:TLR}), (\ref{eq:PLR}), and (\ref{eq:JLR}) (dashed lines), which connect the output (temperature, power, or heat) to the input (the resonator frequency) by multiplying it with a response function. The amplitude of the output is given by the absolute value of the response function times the input, while the time delay is determined by the complex phase of the response function. For example, at high frequencies, the first fraction in Eq.~(\ref{eq:TLR}) approaches one, and the temperature gets synchronized with the drive as in Eq.~(\ref{eq:Wtemp}). With a large driving amplitude, the results in Fig.~\ref{fig:fig4} do not follow the linear-response prediction. By contrast, with a small driving amplitude, we find good agreement between them. In particular, according to Eqs.~(\ref{eq:InputLR})~and~(\ref{eq:OutputLR}), the linear-response results should describe an ellipse in a parametric phase-space plot as in the lower row of Fig.~\ref{fig:fig4}. \revision{The frequency-dependent response functions found above reveal that the system behaves as a high-pass filter for temperature and heat, with the coupling $\gamma$ setting the bandwidth. This provides a direct way to engineer dynamical susceptibilities of quantum thermal devices.}

\section{Fluctuations}
\label{sec:fluc}

We now investigate the fluctuations in the driven resonator beyond the average values considered so far. In particular, we are interested in the statistics of quantum jumps, which one may think of as the exchange of photons between the resonator and the environment. \revision{While full counting statistics of quantum resonators have been investigated in stationary settings~\cite{Clerk:2007,Hofer:2016,Brange2019,Menczel:2021,Portugal:2023,Kansanen:2025}, their behavior under time-dependent driving remains largely unexplored. Here, we show that temporal modulation leads to dynamically evolving, strongly non-Gaussian statistics, which cannot be inferred from steady-state results.}
Specifically, we consider the net transfer of photons, which may be detected using calorimetric measurements~\cite{Pekola:2013}. However, our approach can  be adapted to other physical situations, for example, if only the emitted photons can be detected but not those that are absorbed.

To describe the photon counting statistics, we resolve the density matrix with respect to the number of transferred photons $m$ and write it as $\hat \rho(m,t)$~\cite{Plenio:1998,Landi:2024}. From the $m$-resolved density matrices, we obtain the statistics of transferred photons during the time span~$[0,t]$ as 
\begin{equation}
p(m,t)=\mathrm{tr}\{\hat \rho(m,t)\}.
\end{equation}
The equations of motion for $\hat \rho(m)$ follow from Eq.~(\ref{eq:LindbladME}),
\begin{equation}
            \partial_t \hat \rho(m) =\mathcal{L}_0\hat \rho(m)
            +\mathcal{J}_e\hat\rho(m-1)
            +\mathcal{J}_a\hat\rho(m+1),
\label{eq:GMEn}
\end{equation}
having introduced the jump operators for the emission, 
\begin{equation}
\mathcal{J}_e\hat\rho=\gamma \big(1+n_B(\omega_0) \big)\hat a \hat \rho\hat a^\dagger,
\end{equation}
and the absorption of photons,
\begin{equation}
\mathcal{J}_a\hat\rho=\gamma n_B(\omega_0) \hat a^\dagger \hat \rho\hat a.
\end{equation}
We have also defined the time-evolution operator without these quantum jumps, $\mathcal{L}_0=\mathcal{L}-\mathcal{J}_e-\mathcal{J}_a$, where $\mathcal{L}$ is the full Lindblad operator defined in Eq.~(\ref{eq:LindbladME}).

To solve the coupled equations for $\hat \rho(m)$, we introduce the counting field $s$ by defining the density matrix
\begin{equation}
\hat \rho(s) =  \sum_m \hat \rho(m) e^{ms},
\end{equation}
whose time-evolution follows from Eq.~(\ref{eq:GMEn}) and reads
\begin{equation}
            \partial_t \hat \rho(s) =[\mathcal{L}_0+e^s\mathcal{J}_e+e^{-s}
            \mathcal{J}_a]\hat\rho(s)=\mathcal{L}(s)\hat\rho(s),
\label{eq:GMEs}
\end{equation}
where $\mathcal{L}(s)$ is known as the tilted Lindblad operator. We note that the tilted Lindblad operator depends on time because of the time-dependent resonator frequency.

\begin{figure*}[!t]
    \centering
\includegraphics[width=\textwidth]{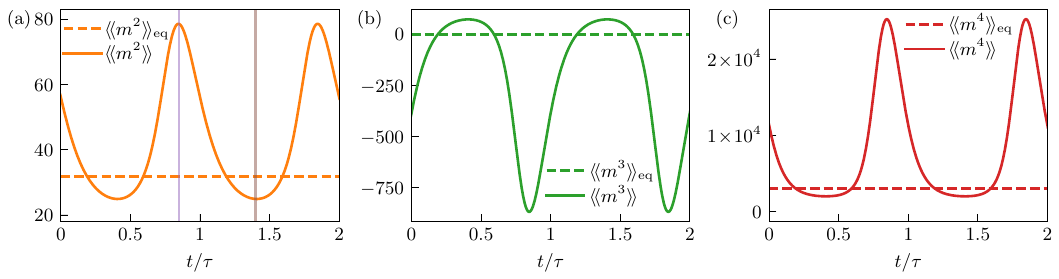} 
    \caption{Second, third, and fourth cumulants. (a,b,c) The solid lines show numerical results for a harmonic drive, while the dashed lines are the cumulants without the drive. The drive reads $\omega_0(t)=\bar\omega_0+\Delta\omega_0\sin(\Omega t)$ with $\Omega = 0.1\bar\omega_0$ and $\Delta \omega_0 = 0.6 \bar \omega_0$. The other parameters are  
    $\gamma = 0.1 \omega_0$ and $k_B T_e = 4 \hbar \bar \omega_0$. The  vertical lines in panel (a) correspond to the distributions in Fig.~\ref{fig:FCS}.}
    \label{fig:cumulants}
\end{figure*}

From the $s$-dependent density matrix, we obtain the moment generating function of the counting statistics as
\begin{equation}
            M(s,t)=\sum_m p(m,t)e^{ms}=\mathrm{tr}\{\hat \rho(s,t)\}.
            \label{eq:MGF}
\end{equation}
The moments of the photon transfers are then obtained as derivatives with respect to the counting field at $s=0$,
\begin{equation}
            \langle m^k\rangle (t) =\partial_s^k M(s,t)|_{s=0}.
\end{equation}
The cumulant generating function is defined as
\begin{equation}
           C(s,t)=\ln M(s,t)=\ln \mathrm{tr}\{\hat \rho(s,t)\},
\end{equation}
and the cumulants similarly follow as
\begin{equation}
           \langle\!\langle m^k\rangle\!\rangle (t) =\partial_s^k C(s,t)|_{s=0}.
\end{equation}
Only the first two cumulants, $\langle\!\langle m\rangle\!\rangle =\langle m\rangle$ and $\langle\!\langle m^2\rangle\!\rangle =\langle m^2\rangle-\langle m\rangle^2$, are nonzero for a Gaussian distribution, while the higher ones, such as the skewness, $\langle\!\langle m^3\rangle\!\rangle$, and the kurtosis, $\langle\!\langle m^4\rangle\!\rangle$, describe non-Gaussian fluctuations.

To find the cumulant generating function, it is convenient to introduce the characteristic function of a state~$\hat \rho$, 
\begin{equation}
    \chi(\lambda ) = \text{tr}\{ \hat D(\lambda) \hat \rho\},
\end{equation}
where $\hat D(\lambda) = e^{\hat a^\dagger \lambda - \hat a \lambda^*}$ is the displacement operator and $\lambda$ is a complex parameter. The characteristic function of the thermal state Eq.~(\ref{eq:effectivethermal}) with average occupation $n$ is
\begin{equation}
\chi(\lambda)= e^{- |\lambda|^2 (n+1/2)}.
\end{equation}
The inclusion of the counting field does not change this functional form, except that $n=n(s)$ becomes $s$-dependent, and the density matrix is no longer normalized, such that the characteristic function becomes 
\begin{equation}
\chi(\lambda, s)= M(s) e^{- |\lambda|^2(n(s)+1/2)},
\end{equation}
where we have used that the prefactor must be the moment generating function, since $M(s)= \chi(0, s)$. Next, we differentiate this expression with respect to time, 
\begin{equation}
    \partial_t\chi (\lambda,s) = \left( \partial_t \ln M(s) - |\lambda|^2 \partial_t n(s) \right) \chi(\lambda,s),
    \label{eq:chi-eom}
\end{equation}
and recognize the cumulant generating function.

The time-evolution of the characteristic function reads 
\begin{equation}
\partial_t\chi (\lambda,s)=\text{tr}\{ \hat D(\lambda) \partial_t\hat \rho(s)\}=\text{tr}\{ \hat D(\lambda) \mathcal{L}(s)\hat \rho(s)\},
\end{equation}
and since the tilted Lindblad operator is quadratic in the creation and annihilation operators, we obtain an equation of motion as in Eq.~(\ref{eq:chi-eom}). We can then identify the time-evolution of the cumulant generating function,
\begin{equation}
\begin{split}
\label{eq:cumulant_net_counting}
    \partial_t C(s) &= \gamma (e^{s} - 1)  n(s)(1 +  n_B(\omega_0))\\
    & +\gamma(e^{-s} - 1) n_B (\omega_0)(1 + n(s)),
\end{split}
\end{equation}
and the $s$-dependent occupation number 
\begin{equation}
\begin{split}
    \partial_t n(s) &= \gamma(e^s-1)n^2(s)(1+n_B(\omega_0))  \\
    &+  \gamma(e^{-s}-1)n_B(\omega_0)(1+n(s))^2\\
    &+\gamma (n_B(\omega_0)- n(s)), 
\end{split}
\label{eq:occupation}
\end{equation}
where the right-hand side in the first equation vanishes for $s=0$, since $C(0)=0$ by definition, while the second equation without the counting field reduces to Eq.~(\ref{eq:n(omega)}). These equations form the basis for our calculations of the fluctuations in the driven resonator.

\section{Cumulants}
\label{sec:cumu}
To find the cumulants of the full counting statistics, we expand the cumulant generating function as
\begin{equation}
    C(s)=\sum_{k=1}^\infty s^k\langle\!\langle m^k\rangle\!\rangle /k!,
\end{equation}
and the $s$-dependent occupation number as
\begin{equation}    n(s)=\sum_{k=0}^\infty s^k n_k/k!.
\end{equation}
Inserting these expansions into Eqs.~(\ref{eq:cumulant_net_counting}) and (\ref{eq:occupation}) and collecting terms to same order in the counting field, we find a hierarchy of coupled equations for each term in the expansions. For the first two terms, we find
\begin{equation}
\begin{split}
    \partial_t \langle\!\langle m\rangle\!\rangle =& \gamma ( n_0-n_B(\omega_0)),\\
    \partial_t \langle\!\langle m^2\rangle\!\rangle =& \gamma ( n_B(\omega_0)(2 n_0 +1 ) +(2 n_1 + n_0)),
\end{split}
\end{equation}
and
\begin{equation}
\begin{split}
    \partial_t n_0 =& \gamma (n_B(\omega_0) - n_0),\\
    \partial_t n_1 =& \gamma ( n^2_0 - n_1) - \gamma n_B(\omega_0)(1 + 2 n_0),
\end{split}
\end{equation}
and similar, but longer, expressions can be found for the higher-order terms. We can then solve these equations numerically, order-by-order, to find each cumulant.

In Fig.~\ref{fig:cumulants}, we show the first few cumulants (full lines) as functions of time, compared to equilibrium results without the drive (dashed lines). The cumulants evolve in response to the harmonic drive with the  distribution being either wider or narrower than the equilibrium result (given by the second cumulant), tilting to the left or to the right (signaled by the sign of the third cumulant), and with light or heavy tails (determined by the sign of the fourth cumulant). The cumulants grow with the driving amplitude, and the exchange of photons is governed by a broad distribution rather than by small fluctuations around the mean. 
\revision{The emergence of large higher-order cumulants indicates that energy exchange occurs via rare but significant photon-transfer events, rather than small Gaussian fluctuations. This intermittency is a hallmark of far-from-equilibrium quantum thermodynamics and may strongly affect the stability and efficiency of nanoscale heat engines.}
To corroborate our findings, we will return to a calculation of the full distribution at the times indicated by vertical lines in Fig.~\ref{fig:cumulants}(a).

\section{Cumulants in linear response}
\label{sec:LR2}

\begin{figure*}
	\centering
	\includegraphics[width=\textwidth]{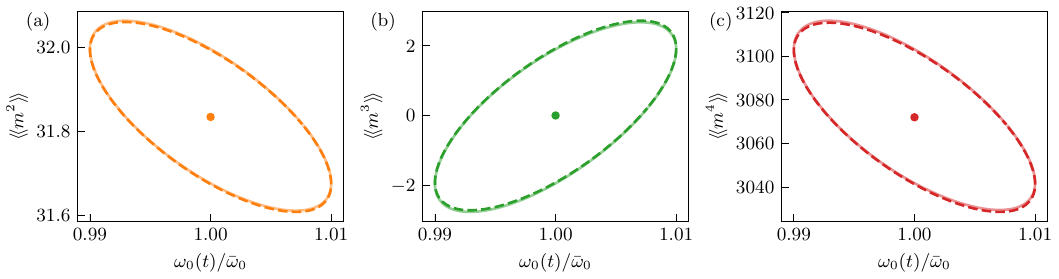}
	\caption{Cumulants in linear response. (a,b,c) Solid lines are numerical results, while the dashed ones show the linear-response expressions. The drive is $\omega_0(t)=\bar\omega_0+\Delta\omega_0\sin(\Omega t)$ with $\Omega = 0.1\bar\omega_0$ and $\Delta \omega_0 = 0.01 \bar \omega_0$. The other parameters are  
		$\gamma = 0.1 \omega_0$ and $k_B T_e = 4 \hbar \bar \omega_0$. The dots correspond to the equilibrium cumulants without the drive.}
	\label{fig:cumulants_LR}
\end{figure*}

Similarly to the temperature, the power, and the heat, we can find the cumulant generating function in linear response. To this end, we write the cumulant generating function and the $s$-dependent occupation number as
\begin{equation}
\begin{split}
C(s,t) &= \bar C(s) + \Delta C(s,t),\\
n(s,t) &= \bar n(s) + \Delta n(s,t),    
\end{split}
\label{eq:CandnLR}
\end{equation}
where
\begin{equation}
\bar C(s) =  \ln(\bar n(s)/\bar n(0))+\ln(\bar n(-s)/\bar n(0))
\label{eq:C(s)_eq}
\end{equation}
and
\begin{equation}
\bar n(s) = n_B(\bar{\omega}_0 + s k_BT_e /\hbar)  
\label{eq:n(s)_eq}
\end{equation}
are those quantities at long times without the drive. Equation~(\ref{eq:C(s)_eq}) is obtained by integrating the time-dependent solution of Eq.~(\ref{eq:occupation}) and taking the limit of long times.  We find Eq.~(\ref{eq:n(s)_eq}) by solving Eq.~(\ref{eq:occupation}) with a constant resonator frequency and the time-derivative set to zero. From Eq.~(\ref{eq:C(s)_eq}) we then see that the odd cumulants vanish in equilibrium, since $\bar C(s)=\bar C(-s)$. By contrast, the second and fourth cumulants are 
\begin{equation}
    \begin{split}
        \langle\!\langle m^2\rangle\!\rangle_{\mathrm{eq}} &= 2 n_B(\bar{\omega}_0) (1 + n_B(\bar{\omega}_0)), \\
        \langle\!\langle m^4\rangle\!\rangle_{\mathrm{eq}} &= \langle\!\langle m^2\rangle\!\rangle_{\mathrm{eq}}(1 + 6 n_B(\bar{\omega}_0) + 6 n_B^2(\bar{\omega}_0)).
    \end{split}
\end{equation}

Inserting Eq.~(\ref{eq:CandnLR}) into Eqs.~(\ref{eq:cumulant_net_counting}) and (\ref{eq:occupation}) and linearizing them around the small changes, we find
\begin{equation}
\begin{split}
\partial_t \Delta C (s,t) & \simeq \gamma  (e^s-1) (1+\bar n(0))\Delta n(s,t)\\
&+ \gamma(e^{-s}-1) \bar n(0)\Delta n(s,t)\\
&+\frac{\gamma \hbar}{k_B T_e} (\bar n(0)-\bar n(s))\Delta\omega_0(t)
\end{split}
\end{equation}
and
\begin{equation}
\partial_t \Delta n (s,t)\simeq  \frac{\gamma \hbar}{k_B T_e}   \Delta\omega_0(t) \partial_s \bar n(s) - \gamma \Delta n (s,t),
\end{equation}
which can be solved in the Fourier domain to finally yield
\begin{equation}
\Delta C(s,\Omega) \simeq \frac{\gamma}{\gamma + i \Omega} (1+\bar n(0)) \bar n(s) (e^s-1) \frac{\hbar \Delta \omega_0(\Omega)}{k_B T_e}.
\end{equation}
For the first cumulant, we now find
\begin{equation}
    \Delta\langle\!\langle   m\rangle\!\rangle(\Omega)  \simeq\frac{\gamma}{\gamma + i \Omega} (1+\bar n(0)) \bar n(0) \frac{\hbar \Delta \omega_0(\Omega)}{k_B T_e}, 
\end{equation}
while, in general, for the $k$'th cumulant, we have
\begin{equation}
    \Delta\langle\!\langle   m^k\rangle\!\rangle(\Omega)\simeq\left[\sum_{l = 0}^{k - 1}  \binom{k}{l} \frac{\bar n^{(l)}(0)}{\bar n(0)}\right]\Delta\langle\!\langle   m\rangle\!\rangle(\Omega),
    \label{eq:LRcumu}
\end{equation}
where $\bar n^{(l)}(0)$ denotes the $l$'th derivative of the function~$\bar n(s)$ in Eq.~(\ref{eq:n(s)_eq}) with respect to $s$ at $s=0$. \revision{Equation~ (\ref{eq:LRcumu}) shows that, even in linear response, higher-order cumulants are not redundant but contain independent information about the system dynamics.}

In Fig.~\ref{fig:cumulants_LR}, we show parametric plots of the first few cumulants and compare numerical results with our linear-response expressions in Eq.~(\ref{eq:LRcumu}). Since the driving amplitude is small, we find good agreement between the two approaches, and we clearly see that the third cumulant is in phase with the drive, while the two others are out of phase. In the linear-response regime, the cumulants are directly proportional to the driving amplitude, and even close to equilibrium, the thermodynamic processes cannot be fully characterized by mean values alone. As such, the fluctuations provide independent and experimentally accessible information about the  resonator dynamics.

\section{Photon distribution}
\label{sec:FCS}

\begin{figure*}
	\centering
	\includegraphics[width=\textwidth]{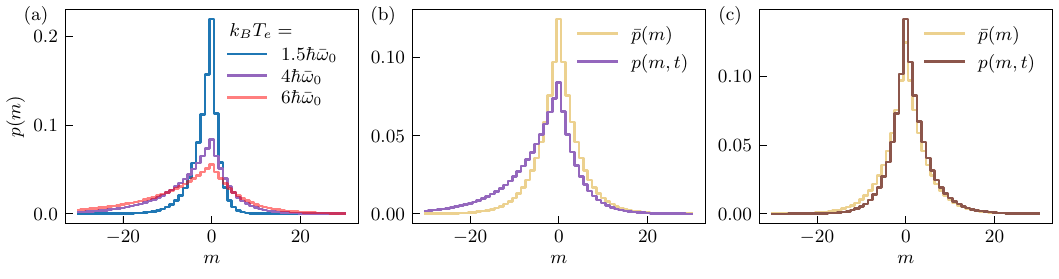}
	\caption{Photon counting statistics. (a) Probability distributions for three different temperatures  the time indicated by a vertical purple line in Fig.~\ref{fig:cumulants}(a). (b,c) Distributions corresponding to the two moments of times indicated by vertical lines in Fig.~\ref{fig:cumulants}(a). The drive reads $\omega_0(t)=\bar\omega_0+\Delta\omega_0\sin(\Omega t)$ with $\Omega = 0.1\bar\omega_0$ and $\Delta \omega_0 = 0.6 \bar \omega_0$. The other parameters are  
		$\gamma = 0.1 \omega_0$ and $k_B T_e = 4 \hbar \bar \omega_0$. For the sake of comparison, we also show the equilibrium distribution without the drive, given by Eq.~(\ref{eq:equidist}).  }
	\label{fig:FCS}
\end{figure*}

We now investigate the full distribution of photons that have been exchanged between the resonator and the thermal reservoir. To begin with, we consider the full counting statistics at long times in the absence of the drive. To this end, we write the moment generating function as 
\begin{equation}
\bar M(s)=\frac{\bar n(s)\bar n(-s)}{\bar n^2(0)},
\end{equation}
having used the equilibrium expression in Eq.~(\ref{eq:C(s)_eq}). We then expand it in the counting field as in Eq.~(\ref{eq:MGF}) and identify the full distribution of transferred photons as 
\begin{equation}
\bar p (m) = e^{-|m|  \hbar\bar \omega_0/k_B T_e} \tanh \left(\hbar \bar \omega_0/2 k_B T_e\right).
\label{eq:equidist}
\end{equation}
For the time-dependent case, we can find the moment generating function by numerically solving Eqs.~(\ref{eq:cumulant_net_counting}) and~(\ref{eq:occupation}) for the cumulant generating function. We then obtain the  probabilities from the inversion formula
\begin{equation}
    p(m, t) = \frac{1}{2 \pi i} \int_{-i\pi}^{i\pi} ds\,  e^{-m s } M(s, t),
\end{equation}
which yields the full distribution of photon transfers.

In Fig.~\ref{fig:FCS}, we show the full distribution of photon exchanges at different times and for different temperatures. In Fig.~\ref{fig:FCS}(a), we see how the distribution grows wider with increasing temperature, as thermal fluctuations become more prominent, and the discreteness of the individual photon transfers becomes less visible. For the sake of comparison, we also show the equilibrium distribution in Eq.~(\ref{eq:equidist}). The shape of the distributions can be directly related to the cumulants in Fig.~\ref{fig:cumulants}. Specifically, in Fig.~\ref{fig:FCS}(b), we see how the distribution becomes wide, left-skewed, and heavy-tailed, corresponding to the results for the cumulants in Fig.~\ref{fig:cumulants}. Similarly, in Fig.~\ref{fig:FCS}(c), the distribution almost has the same width as the equilibrium distribution, and it is slightly right-skewed and light-tailed, consistently with the cumulants in Fig.~\ref{fig:cumulants}.

The results in this section demonstrate that fluctuations provide essential information beyond mean values. In particular, the higher cumulants of the photon distribution reveal clear non-Gaussian features that become larger  outside the linear-response regime. These fluctuations reflect the quantum nature and the stochastic character of energy exchange at the nanoscale.

\section{Conclusions and outlook}
\label{sec:concl}

\revision{We have developed a time-resolved thermodynamic and statistical description of a driven quantum resonator, which combines average energy flows with full counting statistics of photon exchange. Specifically,} we have analyzed the thermodynamic properties of a driven quantum resonator whose temperature is controlled by modulating its natural frequency. By evaluating the work performed by the external drive and the resulting heat exchange with the environment, we have provided a consistent description of energy flows both within linear response and beyond. In addition to average observables, we have found the full distribution of photon exchanges between the resonator and the environment, characterized by its first few cumulants,  highlighting the role of fluctuations in quantum thermodynamic processes.  \revision{Our results show that temporal driving induces qualitatively new fluctuation dynamics, including strongly non-Gaussian and time-dependent cumulants that cannot be captured by steady-state theories. Importantly, we have found that higher-order cumulants provide independent and experimentally accessible information about the system dynamics.} Our framework connects work, heat, and fluctuations in a driven open quantum system, which may function as the working fluid in a quantum heat engine.

\revision{Although we do not consider a full thermodynamic cycle, our results may help optimize quantum heat engines. In particular, the separation of timescales between driving and dissipation, and the presence of  non-Gaussian fluctuations, suggest trade-offs between power, efficiency, and stability. These effects are expected to play an important role in finite-time quantum engines based on resonators.}
An interesting direction for future work is therefore to extend the present analysis to a complete quantum heat engine cycle. Such an analysis would allow for a systematic investigation of performance metrics, including efficiency, power, and their fluctuations, also in regimes far from equilibrium. The combination of frequency modulation with the controlled coupling to several thermal reservoirs could enable the realization and optimization of concrete engine protocols, such as quantum Otto or Stirling cycles, based on resonator platforms.

More broadly, the approach developed here can be applied to a wide range of experimentally relevant systems, including superconducting circuits, optomechanical resonators, and hybrid quantum devices. Incorporating effects such as strong system-environment coupling, non-Markovian dynamics, or quantum measurement backaction represents further promising avenues. Such extensions may improve our understanding of quantum thermodynamics at the nanoscale and help guide the design and optimization of future quantum heat engines and other quantum devices for efficient energy conversion.

\begin{acknowledgements}
We acknowledge the support from the Nokia Industrial Doctoral School in Quantum Technology, the Research Council of Finland through the Finnish Centre of Excellence in Quantum Technology (Grant No.~352925) and the Finnish Quantum Technology Flagship.
\end{acknowledgements}

\section*{DATA AVAILABILITY}
The code used to produce the numerical data presented in this paper is openly available in Ref.~\cite{github}.

%

\end{document}